\documentclass[prl, floatfix, twocolumn]{revtex4}
\usepackage{graphicx}

\newcommand{\eq}[1]{Eq.~\ref{#1}}

\newcommand{\fig}[1]{Fig.~\ref{#1}}

\begin{document}
\title{Hybrid resonant phenomenon in a metamaterial structure with integrated resonant magnetic material}
\author{Jonah N. Gollub${}^1$} 
\author{David R. Smith${}^2$}
\author{Juan D. Baena${}^3$}

\affiliation{${}^1$ Department of Physics, University of California,
San Diego, CA, 92037 USA}
\affiliation{${}^2$ Department of Electrical and Computer
Engineering, Duke University, Durham, NC, 27708 USA}
\affiliation{${}^3$ Department of Physics, National University of Colombia, Bogot\'{a}, Colombia}

\date{\today}

\begin{abstract}
We explore the hybridization of fundamental material resonances with the artificial resonances of metamaterials. A hybrid structure is presented in the waveguide environment that consists of a resonant magnetic material with a characteristic tuneable gyromagnetic response that is integrated into a complementary split ring resonator (CSRR) metamaterial structure. The combined structure exhibits a distinct hybrid resonance in which each natural resonance of the CSRR is split into a lower and upper resonance that straddle the frequency for which the magnetic material's permeability is zero. We provide an analytical understanding of this hybrid resonance and define an effective medium theory for the combined structure that demonstrates good agreement with numerical electromagnetic simulations. The designed structure demonstrates the potential for using a ferrimagnetic or ferromagnetic material as a means of creating a tunable metamaterial structure. 
\end{abstract}

\maketitle

\maketitle

\section{Introduction}
\begin{figure}[t]
\centering
\includegraphics[scale=1]{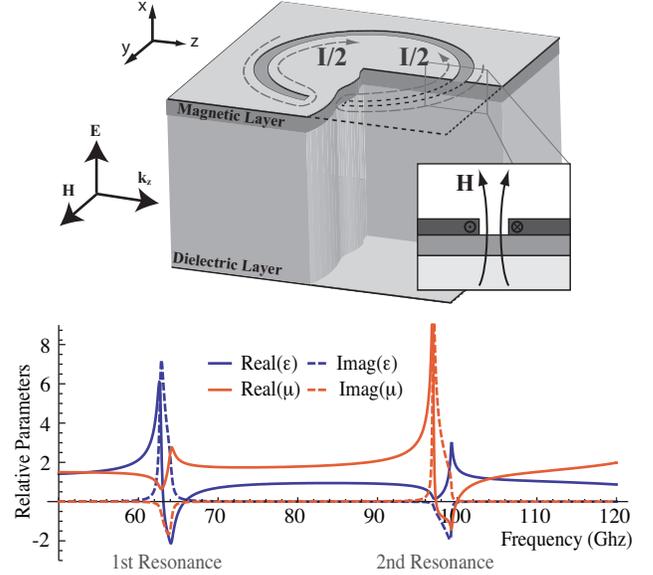}
\caption{A diagram of a unit cell of the CSRR parallel plate waveguide is shown. At resonance, current flows along the edges of the CSRR structure and produces a magnetic field that is approximately perpendicular in the gap region. The extracted permeability and permittivity are shown for the frequency range that includes the first two resonances of the CSRR waveguide (without the influence of the magnetic material).} \label{CSRR_waveguide_diagram}
\end{figure}

Metamaterials encompass a class of artificial electromagnetic media that provide electromagnetic properties beyond those available from natural materials \cite{Veselago1964, Smith2000, Pendry1999SRRs}. Their unique electromagnetic properties are obtained by harnessing the resonant behavior of many periodic and subwavelength composite structures. Unfortunately, the resonant nature of metamaterials also ensures that they are frequency dispersive and limited to a narrow frequency band of operation. One method of bypassing this constraint is to design tunable metamaterials, which though still dispersive, can be tuned to have the desired properties over the frequency range of interest. A tunable metamaterial can be made by placing a material with a tunable electromagnetic parameter, permittivity or permeability, in a region of the metamaterial cell where the local fields are concentrated. Tuning the local fields will in turn affect the bulk electromagnetic response of the metamaterial. This has been demonstrated by tuning the capacitance in the gap region of metamaterials composed of split-ring resonators (SRRs) using ferroelectric materials \cite{Lim2004VaractorTun1DMetaAntenna, Hand2007FerroLoadedSRR}.  In this paper we investigate a tunable metamaterial structure that incorporates a resonant magnetic material, that in contrast to ferroelectric material, is itself frequency dispersive over the range of operation. The two dispersive systems, metamaterial structure and magnetic material, combine to exhibit a distinct hybrid resonance for which we provide an analytical model and demonstrate its good agreement with numerical results.

Gyromagnetic materials have a permeability tensor of the form \cite{Kittel:IntroSolidState},
\begin{equation}
\mu=\mu_0\left(
\begin{array}{ccc}
\mu_1 & i\mu_2& 0 \\
-i\mu_2& \mu_1&0 \\
0&0&1
\end{array}\right),
\end{equation}
where $\mu_1$ and $\mu_2$ are resonant functions of frequency. In this study we consider a simplified ``generic'' resonant magnetic material which is isotropic and does not include the off-diagonal permeability component, $\mu_2$, in order to elucidate the underlying physics of the hybridization. Never-the-less, this structure is suggestive of how ferrimagnetic or ferromagnetic materials might be used to make a tunable metamaterial structure. At the end of this letter we briefly discuss the potential for integrating a real gyromagnetic material into the metamaterial structure.

Metamaterials with artificial magnetic properties have been constructed using SRR structures and exhibit an approximate magnetic Drude-Lorentz response that has been well documented in the literature \cite{Smith2000}. A related resonant structure is called the complimentary split-ring resonator (CSRR) and has the ``complementary" metasurface of a SRR (as shown in \fig{CSRR_waveguide_diagram}). The dual CSRR and SRR structures obey Babinet's principle of equivalence with equal resonant responses \cite{Marques:Babinet_CSRR,Baena2005EqvCSRR_Circ}, but the CSRR is excited by a perpendicular electric field while the SRR structure is excited by a perpendicular magnetic field. 

The CSRR structure is well suited for implementation of the metamaterial concept in a waveguide geometry because the CSRR structures can be etched into the ground plane of a waveguide and excited by an incident TEM wave. Just as for bulk metamaterial structures, it is possible to define effective bulk parameters for a waveguide metamaterial structure \cite{Caloz2001LHMinWaveguide}. By Babinet's equivalence, the CSRRs permittivity is the dual of the SRR structure and is given by
\begin{equation}
\epsilon(\omega)=1-\frac{F\omega^2}{\omega^2-\omega_0^2+i \omega\Gamma_c}, \label{CSRR_drude_permittivity}
\end{equation}
where $F$ is a constant, $\omega_0$ is the resonant angular frequency, $\Gamma_c$ is the dissipation factor, and $\omega$ is the angular frequency \cite{Pendry1999SRRs}. The resonant frequency $\omega_0$ is related to the inductance, $L$, and capacitance, $C$, of the structure by
\begin{equation}
 \omega_0=\frac{1}{\sqrt{LC}}. \label{CSRR_ResonantFreq}
\end{equation}

The inclusion of a thin layer of magnetic material below the CSRR structure as shown in \fig{CSRR_waveguide_diagram} provides a means to influence the the local fields of the CSRR structure. We consider a magnetic material with relative permeability of the form,
\begin{equation}
\mu_r(\omega)=\frac{(\xi M_s+\xi H_0-i\omega\Gamma_m)^2-\omega^2}{(\xi H_0-i\omega\Gamma_m)(\xi M_s+\xi H_0-i\omega\Gamma_m)-\omega^2}\label{MagPermeability}
\end{equation}
where $M_s$ is the magnetic saturation of the material, $H_0$ is the magnetic bias field, $\Gamma_m$ is the damping constant, and $\xi$=$\gamma\mu_0$ is a constant with $\gamma$ defined as the gyromagnetic ratio \cite{Liau1991}. The CSRR resonance and magnetic material resonance interact to produce a hybrid resonance. The mechanism of interaction can be understood by considering the dynamics of the current flow when the CSRR is resonating. An incident TEM wave in the waveguide drives current symmetrically in and out of the CSRR structure along the edges of the gap region as shown in \fig{CSRR_waveguide_diagram}.  The current is concentrated on either side of the gap of the CSRR structure but flows in opposite directions creating a nearly perpendicular magnetic field in the gap and the regions directly above and below the gap. Consequently, determining the inductance of the CSRR structure is analogous to analyzing the {\it magnetic} capacitor model of parallel current sheets filled with some volume fraction $q$ (to be determined) of frequency dependent magnetic material. Exploiting this simple model, the inductance is found to be,
\begin{equation}
L=\mu_0 \left(\frac{\mu_r(\omega)}{\mu_r(\omega)(1-q)+q}\right) g_{\rm geom},\label{CSRR_inductance}
\end{equation}
where $\mu_r(\omega)$ is the relative permeability of the magnetic material and $g_{\rm geom}$ is a constant with units of length that is determined by the geometry of the CSRR structure. If the parallel current approximation were exact then $g_{\rm geom}=(1/2)wd/h$ with $w$ the length of the gap, $d$ the width of the gap, and $h$ the height of the gap (the factor of 1/2 follows from the parallel current flow into the CSRR structure). In general, the geometrical function is more complex but it can be extracted from numerical simulations of the empty CSRR waveguide as we demonstrate below. The capacitance of the CSRR structure follows from the capacitance across the gap (above and below the metal-surface). It is given by, $C=\epsilon_0f_{\rm geom}$, where again $f_{\rm geom}$ is a constant with units of length that is determined by the geometry of the CSRR structure and can be extracted from numerical simulations.

Inserting \eq{CSRR_inductance} into \eq{CSRR_ResonantFreq} provides an equation that can be solved to determine the resonant frequency of the hybrid CSRR/magnetic structure,
\begin{equation}
\omega_0' =\frac{1}{\sqrt{\frac{\mu_r(\omega_0')}{\mu_r(\omega_0')(1-q)+q}}}\,\omega_0. \label{AnalyticFreqSol}
\end{equation} 
where, $\omega_0=1/\sqrt{\epsilon_0\mu_0 f_{\rm geom} g_{\rm geom}}$. We note that \eq{AnalyticFreqSol} reduces to the resonant frequency of the empty structure in the case $\mu_r(\omega)$ is of unity. As previously mentioned, the magnetic field generated in the gap region of the CSRR at resonance is predominately perpendicular (see \fig{CSRR_waveguide_diagram}) and interacts principally with the magnetic material through the $x$-component of the permeability. In fact, simulations (not shown here) have shown that variation of any of the other diagonal permeability components has no effect on the response of the structure. We can insert the permeability \eq{MagPermeability} into \eq{AnalyticFreqSol} and solve to get the new resonant frequency of the hybrid structure.  \eq{AnalyticFreqSol} is transcendental in nature, as a result of the magnetic materials frequency dependence, but it can be solved straightforwardly using numerical methods. 

\begin{figure}[t]
\centering
\includegraphics[scale=1]{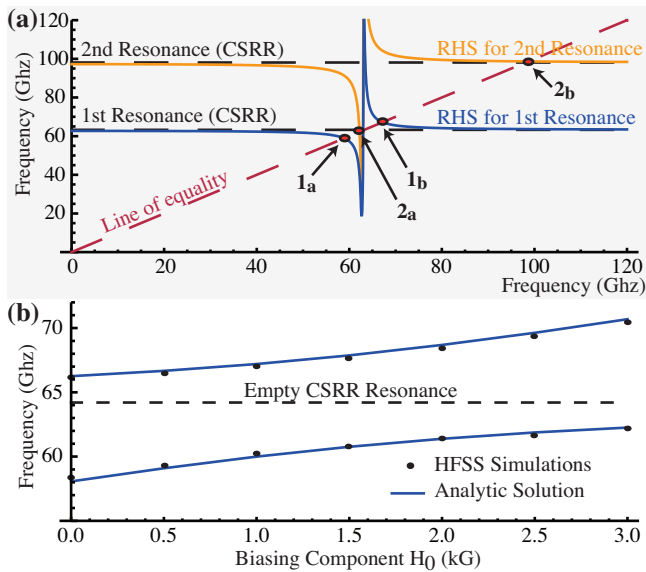}
\caption{(a) The left hand side (line of equality) and right hand side (RHS)  of \eq{AnalyticFreqSol} are plotted for the 1st/2nd CSRR resonance with $H_0=1.5$ kG and characteristic parameters stated in the text. (b) The dispersion curve for the 1st order hybrid mode of the CSRR/magnetic material waveguide is calculated both analytically and through numerical electromagnetic simulations as a function of the biasing field, $H_0$.} \label{hybrid_dispersion_curv}
\end{figure}
In order to understand the dynamics of \eq{AnalyticFreqSol} it is instructive to plot the left and right side of the equation as a function of frequency, as shown in \fig{hybrid_dispersion_curv}(a). Note that the solution to \eq{AnalyticFreqSol} is found at the intersection of these two lines, i.e.\ where the function is self-consistent. The characteristic phenomenon of the hybridization is seen to be a splitting of the CSRR's resonance into a lower, $1_a$,  and upper, $1_b$, hybrid resonance which straddle the frequency at which the magnetic material's permeability is zero (above resonance---not the zero point at resonance).  The hybridization is strongest when the natural resonance of the CSRR structure is near the zero permeability point of the magnetic material. If one increase the biasing field of the magnetic material, such that its zero permeability frequency advances through the CSRR's natural resonance, a characteristic growth and shift of the $1_a$ resonance is observed and then a subsequent decay and shift of the $1_b$ resonance is observed. Hence, tuning the magnetic material effectively tunes the response of the metamaterial structure.  Because a CSRR structure inherently exhibits multiple resonance (the first two resonances of the CSRR structure are shown in \fig{CSRR_waveguide_diagram}) it is not only the fundamental resonance that is hybridized but also the higher order resonances.  However, these higher order resonances are highly damped if they are far away from the magnetic material's zero permeability frequency. In practice this means that if the zero permeability frequency of the magnetic material is aligned with the fundamental CSRR structure resonance, then depending on losses in the system, residual higher order hybrid resonances $\ 2_a,\ {...},\ n_a$ might be found near the fundamental, $1_a$, hybrid resonance as shown in \fig{hybrid_dispersion_curv}(a). In contrast, the $1_b,\ 2_b,\ {...},\ n_a$ resonances remain largely separated. This suggests that the $1_b$ resonance---which is strongly resonant, isolated from the other hybrid resonances, and strongly tunable---maybe the most useful for application.

In order to investigate the hybridization of the CSRR and magnetic material we used Ansoft's commercially available electromagnetic finite element solver (HFSS). HFSS calculates the scattering matrix for a simulated structure via a full wave analysis. It is sufficient to simulate one CSRR unit cell, as shown in \fig{CSRR_waveguide_diagram}, and apply appropriate boundary conditions to reproduce the characteristic response of a parallel plate waveguide structure.  The simulated structure consisted of two parallel perfectly conducting plates and adjoining perpendicular perfect magnetic boundaries. The structure was excited through waveports on either sides of the waveguide with a fundamental TEM mode. The CSRR structure was cut into the top waveguide plate and an exterior volume above the structure was defined in order to accommodate electromagnetic fields emanating above the structure. Perfect magnetic boundaries were defined on the exterior volume's surfaces perpendicular to the propagation mode while perfect electric boundaries were defined on the other surfaces including the top of the extra boundary volume (note that the height of the extra volume is chosen such that on the boundary the electromagnetic fields decay to near zero---and hence its specification is irrelevant, but numerical convergence is faster by defining an electric boundary here).

The geometrical parameters of the CSRR were chosen to have a resonance near 64 Ghz. The unit cell was 1 mm, the ring radius was 0.4 mm, the neck was 0.15 mm wide, and the notch width was 0.035 mm. The spacing between the plates, which determines the strength of the CSRR resonance, was set to 0.4 mm.  The top and bottom waveguide plates were 1 $\mu$m thick and had the conductivity of copper ($\sigma=5.6\times 10^7$). For simplicity, the dielectric between the plates was set to have a dielectric constant of $\epsilon_r=1$. The structure was first simulated without the magnetic layer to determine the Drude-Lorentz fit parameters in \eq{CSRR_drude_permittivity}.   A parameter extraction was performed on the S-parameters \cite{Smith_Retrieval} of the simulated structure to determine the effective permittivity and then a least square fit of the Drude-Lorentz function, \eq{CSRR_drude_permittivity}, was used to determine the resonant frequency, $\omega_0=1/\sqrt{\epsilon_0\mu_0g_{\rm geom}f_{\rm geom}}$; the constant, $F$; and the loss tangent, $\Gamma_c$. In fact, CSRR structures (and metamaterials in general) are not perfect Drude-Lorentz resonators as they exhibit spatial dispersion effects near resonance.  These spatial dispersion effects were incorporated into our fitting procedure \cite{LiuGeneralMetaTheory} to determine the resonant frequency more accurately but then the simpler Drude-lorentz model was used in our analytic analysis with good accuracy.

Once the empty CSRR structure was characterized, numerical simulations incorporating the magnetic material were performed. In the interest of exploring the potential of incorporating high frequency (40-70 Ghz) thin film magnetic materials, such as hexagonal ferrites  \cite{Glass1988:FerriteFilmsReview}, we considered a 1 $\mu$m thick layer of magnetic material with parameters (\eq{MagPermeability}): $\xi=18\ {\rm Ghz\,kG^{-1}}$, $\Gamma_m=(0.70\ {\rm Ghz})/\omega$ and $\xi M_s= 380\ {\rm Ghz}$.  The permeability was calculated in MATLAB and imported into HFSS as a data table of real values (Real[$\mu$]) and loss tangent values (Imag[$\mu$]/Real[$\mu$]). As previously mentioned, it is only the $x$-component of $\mu_r(\omega)$ that contributes to the interaction. Using the fitted parameters extracted from the empty structure we solved the analytic equation, \eq{AnalyticFreqSol}, for the first hybrid mode, $1a$/$1b$, using a numerical root finding method implemented in MATHEMATICA. The magnetic filling fraction, $q$, was determined to have a value of 0.013 through comparison to the HFSS simulations results. A comparison of the analytic dispersion curve calculated from \eq{AnalyticFreqSol} versus the numerical simulations performed with HFSS is shown in \fig{hybrid_dispersion_curv}(b) and demonstrates excellent agreement with variation of the bias field.

\begin{figure}[t]
\centering
\includegraphics[scale=1]{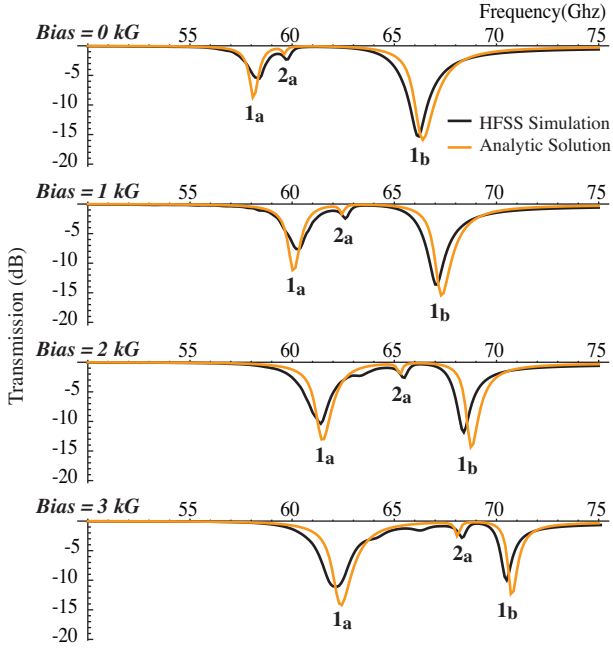}
\caption{The transmission results are shown for the HFSS electromagnetic simulation and the analytic prediction using effective medium theory.} \label{Trans_HFSS_Analytic}
\end{figure}
The transmission through the CSRR/magnetic material waveguide can be calculated analytically by assigning bulk electromagnetic parameters to the waveguide structure.  The effective bulk permittivity of the waveguide structure can be calculated by inputing \eq{AnalyticFreqSol} into \eq{CSRR_drude_permittivity} and then solving the Fresnel equations \cite{Smith_Retrieval} to determine the transmission and reflection of the structure. This was done  for the CSRR/magnetic waveguide structure and compared to HFSS simulations as shown in  \fig{Trans_HFSS_Analytic}. In the HFSS simulations a splitting of the second order CSRR resonance can also be seen. In the analytic expression we were able to reproduce this by applying our technique to the first (electric) and second (magnetic) resonance of the CSRR structure. The overall agreement is seen to correlate well over a range of biasing values.

\begin{figure}[t]
\centering
\includegraphics[scale=1]{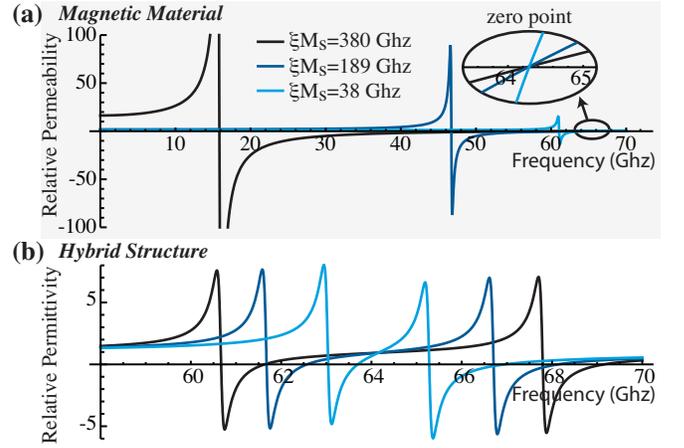}
\caption{(a) The real part of the permeability is shown for several magnetization values of the magnetic material with the bias field chosen to have the same zero permeability frequency. (b) The resulting hybrid permeability is shown.} \label{VaryMagnetization}
\end{figure}
It is remarkable to note that the width and strength of the hybrid resonances is primarily a function of the CSRR's resonant properties.  The properties of the magnetic material's resonance (and filling fraction $q$) predominantly determines the bandwidth over which the hybrid mode interaction exists. In \fig{VaryMagnetization} the permeability of the magnetic material, \eq{MagPermeability}, and the associated effective permeability of the hybrid structure are plotted for various value of magnetization $M_s$. When the biasing field is chosen such that the zero point of the magnetic material correlates to the empty CSRR structure resonance, we see that the relative permeability strength of the hybrid structure is the same but that the splitting of the resonance is smaller for smaller values of $M_s$ values. This suggest that it may be possible to harness narrowly resonant magnetic materials, that cannot be utilized directly, by using this metamaterial approach (though only over a narrow bandwidth).  These potential materials include hexagonal ferrites and even antiferromagnetic materials such as MnF$_2$ \cite{Lui1986:MnF2_Res}.

In summary, the hybrid resonance that results from combining a resonant magnetic material and CSRR structure results in a unique hybrid resonance which can be harnessed to make tunable metamaterial structures. Gyromagnetic materials have a resonant permeability tensor form similar to that considered here but with the added complexity of off-diagonal resonant components.  As a result, the resonant response of the combined structure has a more complicated dependance on the orientation of the biasing direction of the gyromagnetic material with respect to the geometry of the CSRR structure \cite{Marques2000FerriteTL,Horno1990QuasiTEMferrite}. Preliminary work suggest that if we limit ourselves to using thin layers we should expect to see an analogous phenomenon as demonstrated here.  Though gyromagnetic materials have been used directly to make tunable microwave devices \cite{Cramer2000, Astalos1998, Camley:waveguideAnalytic}, using them indirectly in metamaterial structures has the potential advantage of increasing the range of effective material properties while at the same time reducing the amount of the magnetic material needed in the structure and its associated losses.

 This research was supported by U.S. Army Research Office DOA under Grant No. W911NF-04-1-0247. We also acknowledge support from the Air Force Office of Scientific Research through a Multiply University Research Initiative under Contract No. FA9550-06-1-0279.

\end{document}